\documentclass[prl,twocolumn,floats,amssymb]{revtex4}

\input epsf
\usepackage{bm}
\usepackage{amsmath}
\usepackage{graphicx}

\begin{document}

\renewcommand{\bottomfraction}{0.7}
\renewcommand{\topfraction}{0.7}
\renewcommand{\textfraction}{0.2}
\renewcommand{\floatpagefraction}{0.8}
\renewcommand{\thesection}{\arabic{section}}

\addtolength{\topmargin}{10pt}


\newcommand{\eq}{\begin{equation}}
\newcommand{\en}{\end{equation}}
\newcommand{\eqa}{\begin{eqnarray}}
\newcommand{\ena}{\end{eqnarray}}
\newcommand{\eqan}{\begin{eqnarray*}}
\newcommand{\enan}{\end{eqnarray*}}
\newcommand{\spz}{\hspace{0.7cm}}
\newcommand{\lbl}{\label}


\def\Bbb{\mathbb}


\newcommand{\Dslash}{{\slash{\kern -0.5em}\partial}}
\newcommand{\Aslash}{{\slash{\kern -0.5em}A}}

\def\sqr#1#2{{\vcenter{\hrule height.#2pt
     \hbox{\vrule width.#2pt height#1pt \kern#1pt
        \vrule width.#2pt}
     \hrule height.#2pt}}}
\def\smallsquare{\mathchoice\sqr34\sqr34\sqr{2.1}3\sqr{1.5}3}
\def\square{\mathchoice\sqr68\sqr68\sqr{4.2}6\sqr{3.0}6}
 
\def\thinspace{\kern .16667em}
\def\punto{\thinspace .\thinspace}
 
\def\xp{x_{{\kern -.2em}_\perp}}
\def\subp{_{{\kern -.2em}_\perp}}
\def\kperp{k\subp}

\def\derpp#1#2{{\partial #1\over\partial #2}}
\def\derp#1{{\partial~\over\partial #1}}

\def\zbar{\overline{z}}
\def\wbar{\overline{w}}
\def\ez{{{\bf e}}_z}
\def\ezbar{{{\bf e}}_{\zbar}}
\def\vF{ v_{_{\rm F}} }
\def\EF{ E_{_{\rm F}} }
\def\nO{ n_{_{\Omega}} }

\def\pbracks#1{\left\langle #1 \right\rangle}
\def\Expect#1{{\Bbb E}\left( #1 \right)}
\def\Expects#1#2{{\Bbb E}_{#2}\left( #1 \right)}
\def\Irel#1#2{I\kern -0.1em \left(#1 \kern -0.2 em : \kern -0.2 em #2\right)}

\title{How Much Phase Coherence Does a Pseudogap Need?}

\author{Paul~E.~Lammert}\email{lammert@phys.psu.edu}
\affiliation{Department of Physics, 
The Pennsylvania State University,
University Park, PA 16802}
\author{Daniel~S.~Rokhsar}
\affiliation{Department of Physics, 
University of California,
Berkeley, CA 94720,
and the DOE Joint Genome Institute,
Walnut Creek, CA 94598}

\date{Aug. 8, 2001}


\begin{abstract}
It has been suggested that the ``pseudogap'' regime in 
cuprate superconductors, extending up to hudreds of 
degrees into the normal phase, reflects an incoherent
d-wave pairing, with local superconducting 
order coherent over a finite length scale $\xi$, 
insufficient to establish superconductivity.  
We calculate the single-particle spectral density in 
such a state from a minimal phenomenological disordered 
BCS model.  
When the phase-coherence length exceeds the Cooper pair size,
a clear pseudogap appears.
The pseudogap regime, however, is found only over a relatively
narrow range of phase stiffnesses, hence is not expected to
extend more than about 20\% above $T_c$.  

\end{abstract}

\maketitle


It is widely believed that the peculiar normal pseudogap regime 
in underdoped cuprate superconductors (see 
refs. \onlinecite{randeria,RPP} for recent reviews),
holds keys to unraveling the entire problem of cuprate superconductivity.
In this regime, between the superconducting transition 
temperature $T_c$, and $T_*$, possibly hundreds of degrees higher,
spectral density near the putative Fermi surface is supressed
-- in contrast with the familiar behavior of BCS superconductors.  
Evidence of this phenomenon is consistently observed most
clearly in angle-resolved photoemission (ARPES),\cite{loeser,ding}
but also in c-axis tunneling, magnetic susceptibility,
heat capacity, Raman scattering, neutron scattering, 
and NMR measurements. 
The variation of the pseudogap with momentum, strongest near the 
$(\pi,0)$ directions and weak or nonexistent near $(\pi,\pi)$,
mirrors that of the full $d_{x^2-y^2}$ superconducting gap.

One potential explanation of this behavior, as suggested by
Kivelson and Emery\cite{emery-kivelson}, is that {\it local\/} 
superconducting correlations (Cooper pairing)
set in at $T_*$, but that long-range phase coherence is not 
established until the temperature drops below $T_c$.  
The plausibility of the idea derives from the facts that phase 
fluctuations generally play a larger role in lower- (here two) dimensional
systems, and the superfluid density in the cuprates is 
very low, making the order parameter phase ``floppy.''
The notion is analogous to a magnetic material in which
local moments form far above the temperature at which they 
become ordered.  Nevertheless, this is not quite the same thing
as pre-formed pairs.
 
\begin{figure}[h]
\centerline{\rotatebox{0}{
\resizebox{80 mm}{!}{\includegraphics{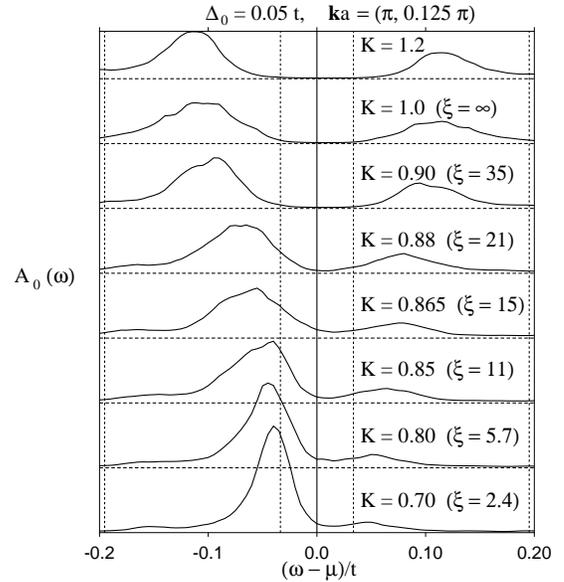}}
 }}
\caption{spectral weight at $((3/24)\pi,\pi)$ 
for $\Delta_0 = 0.05 t$.
Individual energy levels are broadened by convolution
with a Lorentzian of FWHM $t/100$.
The correlation length is computed from the 
Kosterlitz-Thouless expression 
$\xi/a = 0.15 \exp\left({1.82 \sqrt{t}}\right)$\cite{gupta,olsson},
with $t=K/(1-K)$. 
}
\label{spec2303a}
\end{figure}

Several previous calculations have investigated this 
idea\cite{engelbrecht,franz,dorsey,schafer}.
With the exception of reference \onlinecite{franz},
which studies effects of uniform superflow fluctuations,
these are essentially diagrammatic approaches, 
with the attendent need for various sorts of 
truncations or resummations. 
The purpose of the present work is to get a better
understanding of the possibilities and limitations
of the incoherent pairing scenario in a BCS-like framework
with as few additional uncontrolled approximations or dubious
assumptions as possible.  
We do that by investigating a minimal phenomenological model 
permitting essentially exact 
(albeit numerical and statistical) solution. 
This model has the following ingredients.

\noindent (1)
Electrons propagate according to a tight-binding model 
in the presence of a disordered pair order parameter with 
local $d_{x^2-y^2}$ symmetry:
\begin{equation}
H[\Delta] = -\sum_{ij} T_{ij} c_{i\sigma}^\dagger c_{j\sigma}
- \sum_{ij} \left( \Delta_{ij} c_{i\uparrow}^\dagger c_{j\downarrow}^\dagger
+ {\rm h.c.} \right).
\label{ham}
\end{equation}
The chemical potential has been included in the hopping matrix  
$T_{ij} \equiv t_{ij} + \mu \delta_{ij}$.
We take the ratio of next-nearest-neighbor hopping ($t^\prime$)
to nearest-neighbor ($t$) to be 
$t^\prime/t = -0.35$\cite{nazarenko}, and the chemical potential 
$\mu  = -1.1076 t$, corresponding to a doping of $x=0.15$. 
We estimate $t \approx$ 1/5 eV, roughly 1/4 of the observed 
bandwidth.  

\noindent (2)
The complex pair field $\Delta_{ij}$ ($= \Delta_{ji}$)
couples to singlet pairs spanning nearest-neighbor 
sites $i$ and $j$.  In a $d_{x^2-y^2}$ superconductor, 
the (spatially uniform) pair field $\Delta_{ij}$ is 
$\Delta_0$ on horizontal bonds, and $-\Delta_0$ on 
vertical bonds.

\begin{figure}
\centerline{\rotatebox{0}{
\resizebox{70 mm}{!}{\includegraphics{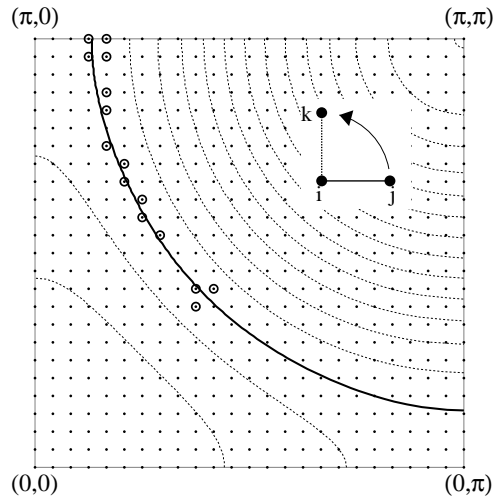}}
 }}
\caption{Upper right-hand quadrant of the Brillouin zone grid
used for numerical computation.  The heavy solid curve indicates
the normal state Fermi surface for the parameters ($t, t', \mu$) reported
in the text.  The dashed curves are normal-state energy contour lines at
a spacing of $t/2$.  Circled points are those for which 
spectral density results are displayed in this paper.  Inset: a pair 
hopping event.}
\label{pox}
\end{figure}

\noindent (3)
In a traditional BCS approach, the gap parameter $\Delta_0$ 
is computed self-consistently taking into account the 
underlying pairing interaction.  
In our phenomenological approach, this linkage is decoupled.
Instead, the pair field $\Delta_{ij} = \Delta_0 e^{i\theta_{ij}}$
in eq. (\ref{ham}) is treated as a quenched random variable, 
chosen from a thermal ensemble.
We assume that $|\Delta_{ij}|$, the magnitude 
of the pairing correlations is a fixed (large) energy scale
$\Delta_0$ equal to the full gap below $T_c$, but that 
the phase suffers thermal fluctuations determined by
the classical XY model
\begin{equation}
\beta H_{XY} = 1.12\, K 
\sum_{\langle ij,ik \rangle} \cos(\theta_{ij} - \theta_{ik}).
\label{xymodel}
\end{equation}
The sum runs over all pairs of nearest-neighbor links, {\em i.e.,} 
links $ij$ and $ik$ related to each other by a $90^\circ$ rotation
about a common site $i$.  The centers of these links form a square
lattice, canted at 45$^\circ$ to the lattice of tight-binding
sites.  
The factor of 1.12 in eq. (\ref{xymodel}) means that $K$ is measured 
in units of the critical coupling, {\it i.e.,\/} the 
Kosterlitz-Thouless-Berezinskii (KTB) transition occurs at 
$K=1$\cite{gupta}.
One may think of the Hamiltonian of Eq. \ref{xymodel} as arising 
from the Hamiltonian 
\begin{equation}
{\cal H}_{pair} = 
g \sum_{i;jk} \Delta_{ij}^* \Delta_{ik},
\label{pairhop}
\end{equation}
which describes the pivoting\cite{pivot} of a Cooper pair from a 
nearest-neighbor link $ik$ to a perpendicular link $ij$
(see inset of Fig. \ref{pox}). 
Positive $g$ encourages local d-wave order. 

\begin{figure}[h]
\centerline{\rotatebox{0}{
\resizebox{!}{80 mm}{
   \includegraphics{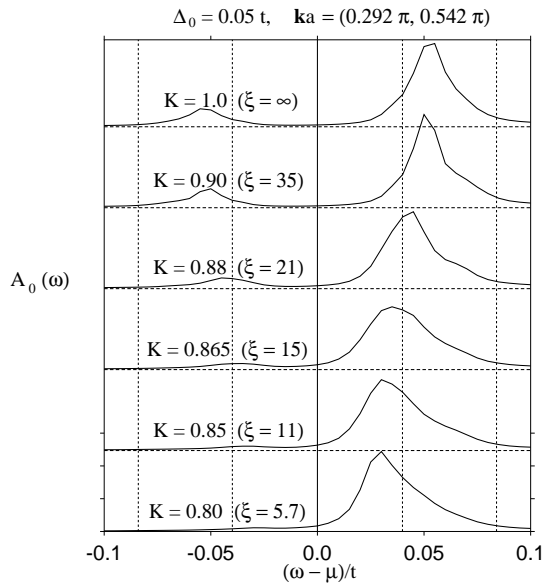}
}}}
\caption{spectral weight at $(({3}/24)\pi,\pi)$ 
for $\Delta_0 = 0.15 t$.
}
\label{spec2303c}
\end{figure}

\begin{figure}
\centerline{\rotatebox{0}{
\resizebox{80 mm}{!}{\includegraphics{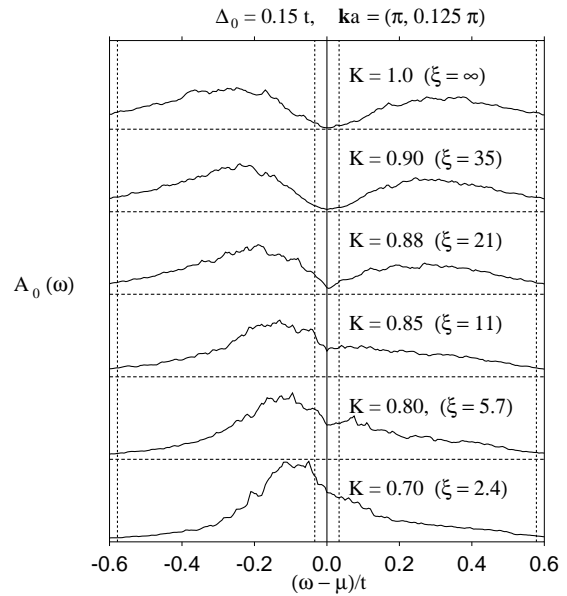}}
 }}
\caption{Spectral weight at $(7,13)\frac{\pi}{24}$ 
for $\Delta_0 = 0.05 t$.
Note the curious feature that the peak is {\em below\/} 
the $\Delta \equiv 0$ value for smaller values of $K$.
The average, however, is closer because of the long
high-energy tail.
}
\label{spec1307a}
\end{figure}

Assumptions 1--3 above specify our model. 
In contrast to BCS theory, a distinction is drawn between 
the local pairing strength $\Delta_0$ and the phase stiffness $K$; 
we take them as independent phenomenological parameters.   
A genuinely two-dimensional system of this sort 
develops only algebraic order for $K > 1$.  
In a real material, coupling between planes causes crossover 
to three-dimensional behavior and true long-range order.
For $K < 1$, there is no superconductivity, but local 
superconductor-like correlations.  
The primary questions are: for what range of phase stiffness $K$ 
does the local pairing manifests itself above the KTB 
transition, and, to how high a temperature is the picture 
able to hold together?
  
{\bf Methods.}
Given a pair field $\Delta_{ij}$, we solve numerically 
for the ground state and quasiparticles by a 
suitable Bogoliubov transformation:
\begin{equation}
\Gamma_m \equiv
\left(\begin{array}{c} 
\gamma_\uparrow \\ 
\gamma_\downarrow^\dagger \end{array} \right)_m
= \left( \begin{array}{cc} U & V \\ -V^* & U^* \end{array} \right)_{mi}
\left(\begin{array}{c} 
c_\uparrow \\
c_\downarrow^\dagger
\end{array} \right)_i,
\end{equation}
where the index $i$ labelling lattice sites is summed over on the 
right-hand side.  Since the $\Gamma$ operators destroy quasiparticles,
we have the Bogoliubov-de Gennes equation
\begin{equation}
\begin{pmatrix} 
-T & {\Delta}^* \\ {\Delta} & T  
\end{pmatrix} 
\begin{pmatrix} 
U_m^T \\ V_m^T  
\end{pmatrix} 
= -E_m
\begin{pmatrix} 
U_m^T \\ V_m^T  
\end{pmatrix}, 
\label{BdG}
\end{equation}
where $U^T_m$ and $V^T_m$ are column vectors to be determined.

With Bogoliubov quasiparticle states in hand, we calculate 
the experimentally relevant particle and hole spectral densities
$A^+(\bm{k} \omega) \equiv 
\langle c({\bm k}) \delta(H-\omega) c({\bm k})^\dagger \rangle$
and
$A^-(\bm{k} \omega) \equiv 
\langle c({\bm k}) \delta(H-\omega) c({\bm k})^\dagger \rangle$.
Since the basic assumption is that the crucial thermal
effects are fluctuations of the order parameter phase,
we compute the quenched average spectral densities at 
zero ``quasiparticle temperature''\cite{qpT}
over the distribution of $\{\Delta_{ij}\}$:
\begin{eqnarray}
A_0^+(\bm{k} \omega) & = & 
\left\langle
\sum_m |U_{m\bm{k}}|^2 \delta(\omega-E_m) \right\rangle
 \nonumber \\
A_0^-(\bm{k} \omega) & = & 
\left\langle
\sum_m |V_{m\bm{k}}|^2 \delta(\omega-E_m)
\right\rangle. 
\end{eqnarray}

Representative full spectral densities $A_0({\bm k},\omega)$
[equal to $A_0^-({\bm k},-\omega)$ for $\omega < 0$ and
to $A_0^+(\bm{k},\omega)$ for $\omega > 0$] over a range
of $K$ are displayed in Figures \ref{spec2303a} and \ref{spec2303c} 
for a point near the Fermi surface in the $(\pi,0)$ direction
with $\Delta_0/t = 0.05$ and 0.15, respectively. 
Results for a Fermi surface point much closer to the 
$(\pi,\pi)$ (nodal) direction is shown in Figure \ref{spec1307a}.  
These plots are the result of Monte Carlo sampling of
the pairing field ensemble for a $48\times 48$ site lattice,
with averaging over three to five disorder realizations and
crystallographically equivalent $\bm{k}$ points.
In each figure, the inner dashed lines indicate plus
or minus the tight-binding energy $\varepsilon(\bm{k})$, 
and the outer dashed lines are the quasiparticle energies
$E_0(\bm{k})=\pm \sqrt{\varepsilon(\bm{k})^2 + \Delta_0(\bm{k})^2}$
for a perfectly ordered pairing field with the
specified $\Delta_0$ ({\it i.e.,} $K=\infty$). 

What do these results tell us?
First, the single-particle spectral function exhibits a pseudogap
even for fairly short-ranged phase coherence.
Roughly speaking, the phase correlation length, $\xi$, is
the distance over which Cooper pairs propagate coherently,
or the characteristic vortex-antivortex separation. 
The figures show a pseudogap developing by about 
$\xi = 15 a$ for $\Delta_0 = 0.05t$.
For this pairing magnitude, the Cooper pair size, estimated
according to $\xi_0 = \hbar \vF/\Delta_0(\bm{k})$, 
is of order $8a$.
(For this estimate, we use Fermi velocity $v_F$ and full gap at 
the Fermi surface near the $(\pi, 0)$ direction.  Other choices
give an estimate a few times larger.) 

Second, the onset of the pseudogap with $K$ is relatively rapid near
the KTB critical point.  The $(\pi,0)$ gap achieves nearly
half its full value by the KTB transition, continuing to
increase slowly with increasing $K$.  The variation of the 
pseudogap along the Fermi surface closely resembles
the pure $d_{x^2-y^2}$ form.  
If we invert the usual BCS relationship between energy gap 
and pairing amplitude, using the peak position of the spectral 
density as a surrogate for the
BCS quasiparticle energy, we can extract an effective ``gap parameter''
$\Delta_{\rm eff}({\bm k}) \equiv 
\sqrt{E_{\rm peak}({\bm k})^2 - \varepsilon({\bm k})^2}$.
At $K = 0.9$, and for both $\Delta_0/t = 0.05$ and $\Delta_0/t = 0.10$,
the ratio $\Delta_{\rm eff}/\Delta_0(\bm{k})$ of effective gap
to full gap is between 0.41 and 0.63 everywhere on the Fermi
surface (with most points being close to 0.5).

Third, spectral peaks are broad with widths roughly linear in the 
pairing magnitude $\Delta_0$.  
As shown in Figure \ref{spec2303c}, this broadening can wash out 
the clear pseudogap, though it does not affect the {\em shift\/} 
of the peaks.  The widths do not monotonically
decrease as $K$ is increased, but are greatest near $K=1$.

Fourth, although the pseudogap appears at a modest phase correlation 
length (comparable to a few Cooper pair radii), the most natural
interpretation of this in terms of temperature is quite discouraging 
if the aim is to explain experimentally observed pseudogaps.
As stressed earlier, $K$ is a phenomenological parameter, 
but it should probably be close to $|\Delta_0|^2 n_s^0/m$,
where $n_s^0$ is the {\em bare\/} superfluid density.
In that case, $K$ is roughly proportional to $1/T$, and
our results imply $T_*/T_c \lesssim 1.2$.  
The collapse of the pseudogap in our model corresponds to
$\xi \sim \xi_0 = \hbar \vF/\Delta_0$, the Cooper pair size.  
If a psuedogap is to be maintained to high temperature, something 
must halt the decrease of $\xi$ at that point.  In other 
words, the order parameter phase must remain stiff at
wavelengths comparable to $\xi_0$.  This issue is not addressed
by our phenomenological model.

\begin{figure}[h]
\centerline{\rotatebox{0}{
\resizebox{80 mm}{!}{\includegraphics{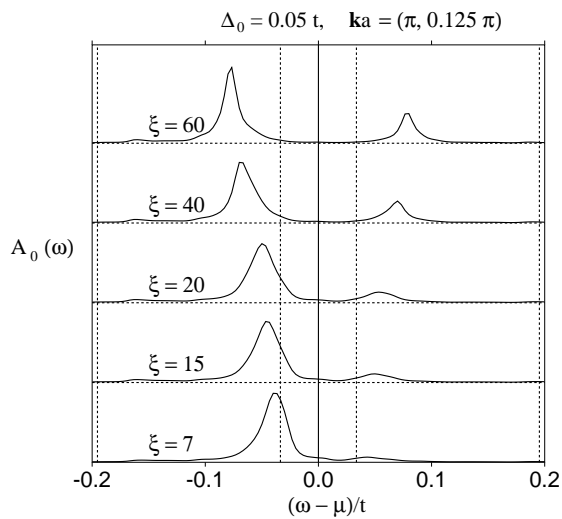}}
 }}
\caption{Spectral density at $(3,23)\frac{\pi}{24}$ 
for $\Delta_0 = 0.05 t$ and several values of 
the phase correlation length $\xi$ (lattice-spacing units), 
calculated by the self-consistent Green function method.
}
\label{selfcon}
\end{figure}

It is interesting to see how well the numerical results 
can be reproduced by a standard diagrammatic approximation,
using a self-consistent Green function.
This approximation is similar to that used by Kwon and 
Dorsey\cite{dorsey}, and is represented by
$$
\epsfxsize 80 mm
\epsfbox{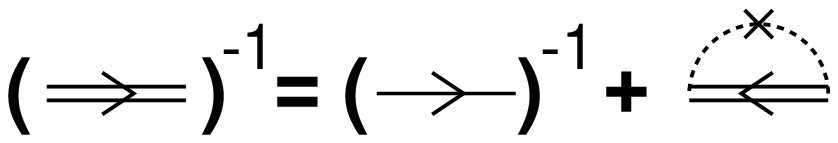}
\label{diagram eqn}
$$
Equivalently, the self-energy is given by
\begin{equation}
\Sigma[G](\bm{k},\omega) = \int \langle \, |\Delta(\bm{q})|^2 \rangle 
G(-\omega,\bm{q}-\bm{k}) \, d^2\bm{q}.
\label{self-energy}
\end{equation}
We solve this iteratively, including $\langle |\Delta(\bm{0})|^2 \rangle$
only in the final iteration.
For $K < 1$, the pair field correlator is accurately approximated by
$
\pbracks{|\Delta({\bm q})|^2}^{-1} \propto \left[\xi^{-2} + 
2 \sum_{{\bm e}}(1 - \cos ({\bm q}\cdot{\bm e})) \right]$,
with a proportionality constant fixed by 
the sum rule $\sum |\Delta({\bm q})|^2  = 2N \Delta_0^2$
($N$ is the number of sites in the lattice).
Spectral densities calculated {\it via\/} this method
(Fig. \ref{selfcon})  also show
the development of a pseudogap at values of $K$ in close 
agreement with the exact results.  However, the peaks are
narrower, by a factor of as much as two.

In summary, we have shown, using a simple phenomenological
model, that a single-particle pseudogap, {\it per se\/},
is natural in the presence of order parameter phase fluctuations.  
In this context, it is unclear what mechanism could maintain short 
range phase stiffness up to the experimentally observed $T_*$.
On the other hand, recent microwave conductivity measurements\cite{orenstein} 
suggest that the {\em bare\/} superfluid density vanishes at a 
temperature comparable to the most natural pseudogap collapse 
temperature found here, about 20\% above T$_c$.
This is also consistent with other previous theoretical 
findings\cite{schafer,dorsey,engelbrecht}.

\begin{acknowledgements}
We thank Steve Kivelson, Joe Orenstein, and Seamus Davis 
for helpful and interesting discussions.  PEL thanks Vincent
Crespi for questions and the David and Lucile Packard Foundation
for financial support.  DSR acknowledges a grant from the UC 
Berkeley Committee on Research and support from the John
Simon Guggenheim Foundation.
\end{acknowledgements}


\begin{thebibliography}{}
\bibitem{qpT} The spectral densities for nonzero quasiparticle
temperature are readily reconstructed:
$ A^\pm({\bm{k}},\omega) = [1-f(\omega)]A_0^\pm({\bm{k}},\omega)
+f(\omega)A_0^\mp({\bm{k}},-\omega)$.
\end{thebibliography}
\end{document}